\documentclass[a4paper]{article}

\usepackage{INTERSPEECH2022}

\usepackage{multirow}
\usepackage{booktabs}
\usepackage{graphicx}
\usepackage[urlcolor=blue]{hyperref}

\title{Content-Dependent Fine-Grained Speaker Embedding for Zero-Shot Speaker Adaptation in Text-to-Speech Synthesis}
\name{Yixuan Zhou$^{1,\dagger}$\thanks{$^{\dagger}$Work conducted when the first author was intern at Tencent.}, Changhe Song$^1$, Xiang Li$^1$, Luwen Zhang$^1$, Zhiyong Wu$^{1,2,*}$\thanks{$^{*}$Corresponding author.},\\ Yanyao Bian$^{3}$, Dan Su$^{3}$, Helen Meng$^{2}$}
\address{
    $^1$ Shenzhen International Graduate School, Tsinghua University, Shenzhen, China\\
    $^2$ The Chinese University of Hong Kong, Hong Kong SAR, China\\
    $^3$ Tencent AI Lab, Tencent, Shenzhen, China
}
\email{zhouyx20@mails.tsinghua.edu.cn, zywu@sz.tsinghua.edu.cn
}

\begin{document}

\maketitle
\begin{abstract}
Zero-shot speaker adaptation aims to clone an unseen speaker's voice without any adaptation time and parameters. 
Previous researches usually use a speaker encoder to extract a global fixed speaker embedding from reference speech, 
and several attempts have tried variable-length speaker embedding.
However, they neglect to transfer the personal pronunciation characteristics related to phoneme content, leading to
poor speaker similarity in terms of detailed speaking styles and pronunciation habits. 
To improve the ability of the speaker encoder to model personal pronunciation characteristics,
we propose content-dependent fine-grained speaker embedding for zero-shot speaker adaptation. 
The corresponding local content embeddings and speaker embeddings are extracted from a reference speech, respectively.
Instead of modeling the temporal relations, a reference attention module is introduced to model the content relevance between the reference speech and the input text, and to generate the fine-grained speaker embedding for each phoneme encoder output.
The experimental results show that our proposed method can improve speaker similarity of synthesized speeches, especially for unseen speakers.

\end{abstract}
\section{Introduction}
\begin{figure*}[!htb]
	\centering
	\includegraphics[width=0.8\linewidth, height=0.27\linewidth]{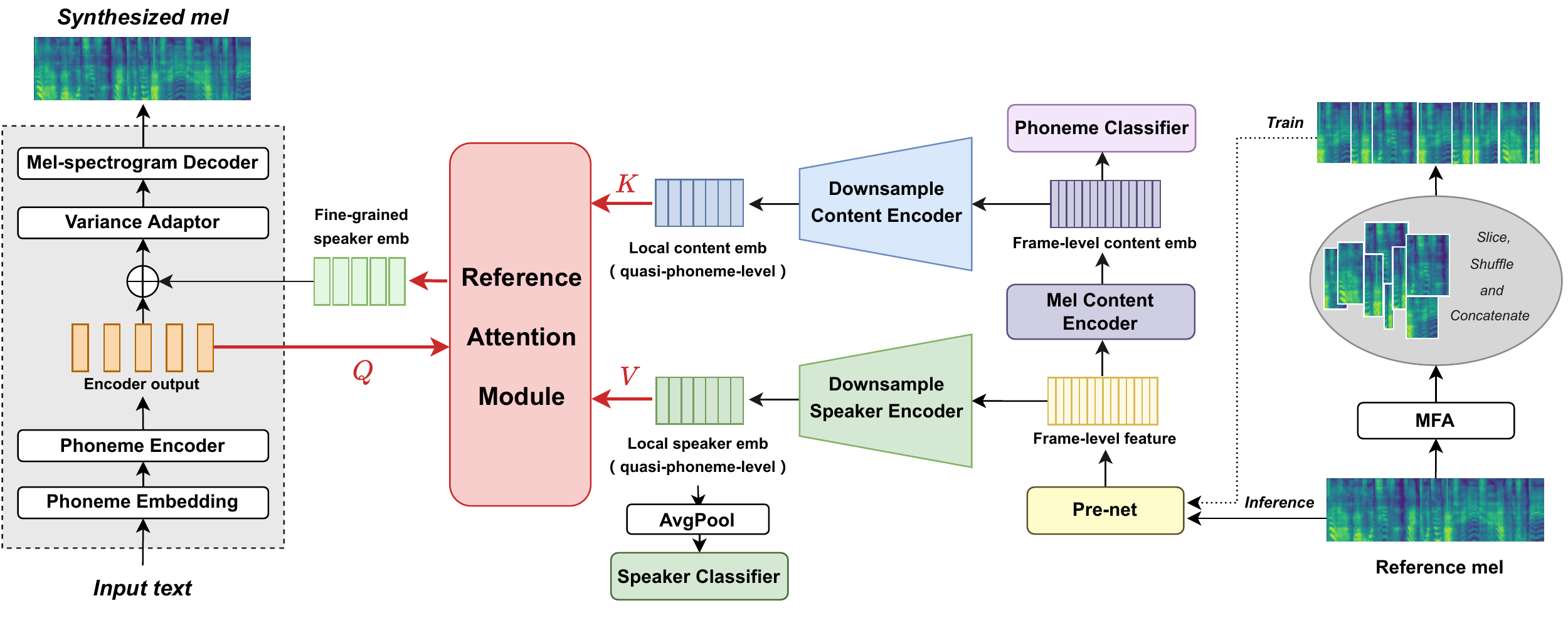}
	\caption{The structure of the proposed model.}
	\label{fig:model_structure}
	\vspace{-1.0em}
\end{figure*}

Neural network-based text-to-speech (TTS), aiming to synthesize intelligible and natural speech from text, has made great progress in recent years \cite{shen2018natural, ping2018deep, ren2020fastspeech}.
These TTS models can synthesize natural human speech with sufficient high-quality training data, for not only single speaker but also multi-speaker scenarios \cite{zhang2019learning, chen2020multispeech}.
But it is too expensive to collect a sufficient amount of speech data for new speakers.
As a result, adapting TTS models to an arbitrary speaker with a few samples (speaker adaptation) is a hot research topic in academia and industry recently \cite{tan2021survey}.

One of the general approaches for speaker adaptation is fine-tuning the whole or part of a well-trained multi-speaker TTS model with a few adaptation data \cite{arik2018neural}.
Some works explore how to better adapt the pre-trained model to the target speaker, such as AdaSpeech series \cite{chen2021adaspeech, yan2021adaspeech2, yan2021adaspeech3}.
These methods are proven to achieve considerable adaptation performance, but with some drawbacks:
(i) certain adaptation time and trainable parameters are required for each new target speaker; 
(ii) voice quality drops quickly when the adaptation data is less than 10 sentences \cite{chen2021adaspeech};
(iii) adaptation performance can be affected by low-quality speeches of the target speaker, resulting in poor intelligibility and prosody of synthesized speech.

To avoid the problems of directly adapting pre-trained models, another line is to leverage a speaker encoder to extract the speaker embedding from reference speech to model speaker identity in TTS.
This approach is also called zero-shot speaker adaptation, since it can clone an unseen speaker's voice by using the speaker embedding only, without any adaptation time and parameters. 
In this connection, it is necessary to explore constructing a better speaker identity representation space to improve the generalization of speaker representation and its adaptability to acoustic models. 
Some researches draw on transfer learning, such as x-vectors from speaker verification tasks \cite{jia2018transfer, cooper2020zero,zhang2021one}.
Others focus on training an encoder network jointly with acoustic models, like using global speaker embeddings (GSEs) \cite{lu2019one} or variational autoencoder (VAE) \cite{hsu2018hierarchical, nguyen2021nvc}.
These methods can clone the overall timbre or speaking style of the reference speech well and make real-time inference for an arbitrary speaker without fine-tuning.

Although representing a speaker's voice with a fixed-length vector is a common idea for zero-shot speaker adaptation, the speaker characteristics of a person actually include not only global timbre information but also some local pronunciation variations.
It is difficult to use a single speaker embedding to describe these local characteristics, leading to poor similarity in terms of detailed speaking styles and pronunciation habits of the target speaker.
Inspired by related works on prosody transfer \cite{klimkov2019finegrained, lee2019robust, li2021towards}, some prior researches try to introduce fine-grained speaker embedding via attention mechanism for capturing more speaker information from speech.  
\cite{Fu2019phoneme} considers phoneme level speaker representations to generate phoneme-dependent speaker embedding by attention.
However, the extraction procedure of phoneme level representations is too complicated and not suitable for zero-shot scenarios.
To make good use of reference speech, 
Attentron \cite{choi2020attentron} proposes an attention-based variable-length embedding method to leverage features near to raw reference speech for better generalization.
However, it only extracts simple reference embeddings without clear meaning and does not show the ability to transfer personal pronunciation characteristics related to phoneme content.

To further improve speaker similarity for zero-shot speaker adaptation,
we extract the corresponding local content embeddings and local speaker embeddings from a reference speech to model personal pronunciation characteristics.
A content-dependent reference attention module is introduced to model the content relevance between the reference speech and the input text, and is used to guide the generation of fine-grained speaker embedding for each phoneme encoder output.
The experiment results show that our proposed method outperforms both two fixed-length speaker embedding methods and a variable-length speaker embedding method based on Attentron in terms of speaker similarity, 
especially for unseen speakers.
The synthesized speeches and experimental analysis demonstrate that our method has the ability to transfer personal pronunciation characteristics related to phoneme content. 
Besides, we investigate the impact of local speaker embeddings with different granularity on the synthesized speech and present the interpretability of our method through visualization.

\section{Methodology}
The model structure of our proposed method is illustrated in Fig.\ref{fig:model_structure}.
We adopt FastSpeech 2 \cite{ren2020fastspeech} as the model backbone, and design several encoders with a reference attention module to obtain content-dependent fine-grained speaker embeddings.
The encoders are used to extract local content and speaker embeddings from the reference mel-spectrograms.
The extracted content and speaker embeddings are then
passed to the reference attention module as the keys and values, while phoneme encoder outputs from FastSpeech 2 are used as queries.
The outputs of the reference attention are then added to the phoneme encoder outputs and passed to the variance adaptor of FastSpeech 2, to generate speech with the same voice as the reference utterance.

\subsection{Extracting local content and speaker embeddings}
To model and transfer personal pronunciation characteristics,
we first extract the corresponding local content embeddings and speaker embeddings from the reference mel-spectrograms.

As shown in Fig.\ref{fig:model_structure}, the reference mel-spectrograms are first passed to a pre-net which consists of two 1-D convolutional layers containing 512 filters with shape $5\times1$.
The frame-level features from the pre-net are encoded by a mel content encoder composed of 4 feed-forward Transformer blocks to get frame-level content embeddings.
For constraining the mel content encoder to encode content information,
a phoneme classifier is introduced to predict the frame-level phoneme labels from the outputs of the mel content encoder.
Then the frame-level content embeddings are passed to the downsample content encoder, meanwhile, the frame-level features are passed to the downsample speaker encoder.
Both two downsample encoders are made up of 4 1-D convolutional layers and a 256-dim fully-connected output layer.
The 4 convolutions contain 128, 256, 512, 512 filters with shape $3\times1$ respectively, each followed by an average pooling layer with kernel size 2.
That is, the temporal resolution is reduced 16 times, which can be regarded as quasi-phoneme level inspired by \cite{li2021towards}. 
All the convolutional layers are followed by ReLU activation and batch normalization \cite{ioffe2015batch}, while the output layer is followed by Tanh activation.
To introduce speaker information, an average pooling layer is used to summarize the local speaker embeddings across time followed by a speaker classifier.
Local content embeddings and local speaker embeddings are obtained from two downsample encoders respectively.
Due to the same local segment input and the same downsampling scale encoding structure, 
they are exactly one-to-one correspondence in the speech.
Therefore, each local speaker embedding can be considered as carrying fine-grained speaker characteristics related to phoneme content.

\subsection{Content-dependent reference attention module}

The speaker characteristics of a person include not only global timbre information but also some local pronunciation variations.
These local variations contain different pronunciation patterns
affected by one's pronunciation habit, which work on a small scale like phoneme level.
For example, there is a difference between a person's pronunciation of ``/\ae/" and his pronunciation of ``/i:/".
Thus, more accurate fine-grained speaker embedding shall be applied to a certain phoneme in text.

The content of the reference speech and input text is different in phoneme permutation and combination during synthesis.
To make better use of local speaker embeddings extracted from reference speech, a content-dependent reference attention module is introduced to obtain the appropriate fine-grained speaker embeddings inspired by \cite{li2021towards, choi2020attentron}.

We adopt scaled dot-product attention \cite{vaswani2017attention} as the reference attention module.
The current phoneme encoder output is used as the query, while all the local content embeddings from reference speech are used as keys.
The relevance between them is used to guide the selection of fine-grained speaker embeddings, which means the local speaker embeddings are values.
In this manner, the fine-grained speaker embedding sequence generated by the reference attention has the same length as the phoneme embedding sequence.

\subsection{Preprocessing operations in the training stage}

The fine-grained characteristics of a speaker are very diverse, for example, the style and pronunciation details are not exactly the same even if one speaker says a sentence twice.
Regarding this, the reference and target utterance had better be consistent in the training stage so that the model can learn correct content relevance and transfer meaningful fine-grained speaker embeddings.
However, the reference attention module easily learns the temporal alignment between reference speech and input text in the previous trial \cite{li2021towards}.
Such fine-grained embedding sequence is more about modeling prosodic trends in time, which is however unsuitable for the input text whose content is different from the reference speech, and will result in strange prosody or poor intelligibility of the synthesized speech in this situation.

To make the model focus more on content relevance rather than simple temporal alignment between reference speech and input text, 
we introduce some preprocessing operations in the training stage.
The mel-spectrogram of a reference utterance is first labeled with frame-level phoneme tags by forced alignment \cite{mcauliffe2017montreal} and divided into fragments by phoneme boundaries.
These fragments corresponding to phonemes are randomly shuffled and concatenated to form a new reference mel-spectrogram.
In this way, the temporal consistency between the paired text and the reference speech is eliminated, and the basic content information of the speech also can be preserved.
The shuffled frame-level phoneme tag sequence is sent to the phoneme classifier as the ground truth for calculating the cross-entropy phoneme classification loss that is added to the total loss.
\section{Experiments}

\subsection{Training setup}
All the models are trained on AISHELL-3 \cite{shi2020aishell}, which is an open-source multi-speaker Mandarin speech corpus containing 85 hours of recordings spoken by 218 native Chinese speakers.
To evaluate the performance on unseen speakers, 8 speakers (4 male and 4 female) are selected 
as the test set.
For the remaining 210 speakers, 95\% of the utterances are used for training and 5\% are used for validation.
Waveforms are transformed to 80-dim mel-sepctrograms with 22.05kHz sampling rate.
The frame size is 1024 and the hop size is 256.
Raw text is converted to phoneme sequence composed of Pinyin initials and tonal-finals by a Chinese grapheme-to-phoneme conversion toolkit\footnote{Pypinyin: \href{https://pypi.org/project/pypinyin}{https://pypi.org/project/pypinyin}}.
We train all the models for 250K iterations with a batch size of 16 on an NVIDIA P40 GPU.
The Adam optimizer is adopted with $\beta_1=0.9$, $\beta_2=0.98$, $\epsilon=10^{-9}$.
Warm-up strategy is employed before 4000 iterations.
A well-trained HiFi-GAN \cite{kong2020hifi} is used as the neural vocoder to generate waveforms.

\subsection{Compared methods}
We compare the proposed content-dependent fine-grained speaker embedding (\textbf{CDFSE}) approach with two typical fixed-length speaker embedding methods and a variable-length embedding method based on Attentron.
These three methods are also implemented based on FastSpeech 2\footnote{Implemented based on: \href{https://github.com/ming024/FastSpeech2}{ https://github.com/ming024/FastSpeech2}}.  

\textbf{GSE} 
Global speaker embedding (GSE) uses a bank of base vectors and multi-head attention to represent the global speaker embedding from reference speech unsupervisedly. 
The implementation is consistent with the original method \cite{lu2019one}.
We also try more base vectors but observe no difference in performance.

\textbf{CLS} 
The speaker classifier (CLS) is a kind of supervised speaker encoder based on multi-task learning or transfer learning \cite{arik2018neural, jia2018transfer, cooper2020zero}.
To compare with the proposed, we use the same speaker encoder as shown in Fig.\ref{fig:model_structure}.
The utterance-level speaker embedding generated by the average pooling layer is replicated to phoneme level and added to the phoneme encoder outputs.

\textbf{Attentron*} 
Attentron proposes an attention-based variable-length embedding method to leverage features near to raw reference speech for better generalization.
It is originally implemented based on Tacotron 2 \cite{shen2018natural}, 
consisted of a coarse-grained encoder and a fine-grained encoder with attention mechanism, which extracts both utterance-level and frame-level embeddings from reference speech. 
To compare with the proposed, we use Attentron (1-1) mode (details in \cite{choi2020attentron}) and adapt its major implementation to FastSpeech 2 framework, named as Attentron*.
The several adjustments are to keep the main structure of the acoustic model unchanged, including:
i) The utterance-level embedding from the coarse-grained encoder is added to encoder output rather than concatenated; ii) The outputs of FastSpeech 2 decoder (before the mel linear layer) are directly used as the queries for attention mechanism to generate frame-level embeddings instead of the autoregressive way in Attentron.

\subsection{Subjective evaluation}

By following \cite{choi2020attentron}, 
we employ two mean opinion score (MOS) tests to evaluate the naturalness and speaker similarity of the synthesized speeches\footnote{Speech samples, detailed figures and open-sourced codes are available at:
\href{https://thuhcsi.github.io/interspeech2022-cdfse-tts}{https://thuhcsi.github.io/interspeech2022-cdfse-tts}}.
8 unseen speakers from the test set and 6 seen speakers randomly selected from the training set are used as reference voices.
The text sentences are from the test set, varying in length and content.
For each speaker, only one utterance is used as the reference speech to guide speech synthesis.
15 native Chinese speakers serves as subjects to take part in the evaluation and rate on a scale from 1 to 5 with 1 point interval. 

\begin{table}[th]
\renewcommand{\arraystretch}{1}
  \vspace{-0.5em}
  \caption{The MOS on naturalness and SMOS (similarity MOS) on speaker similarity with 95\% confidence intervals.}
  \label{tab:mos}
  \centering
  \resizebox{0.99\columnwidth}{!}{
  \begin{tabular}{l|ccc} 
    \toprule
    \textbf{Metric} & \textbf{Model} &\textbf{seen speakers} & \textbf{unseen speakers}\\
    \midrule
    \multirow{4}{*}{MOS} & GSE & $3.50\pm0.16$ & $3.56\pm0.12$ ~~~               \\
    & CLS & $3.51\pm0.14$ & $3.53\pm0.11$ ~~~               \\
    & Attentron* & $3.63\pm0.16$ & $3.57\pm0.13$ ~~~               \\
    & CDFSE & $3.59\pm0.17$ & $3.54\pm0.12$ ~~~ \\
    \bottomrule
    \midrule
    \multirow{4}{*}{SMOS}& GSE & $3.89\pm0.14$ & $3.08\pm0.14$ ~~~               \\
    & CLS & $3.79\pm0.16$ & $3.12\pm0.14$ ~~~               \\
    & Attentron* & $4.04\pm0.17$ & $3.29\pm0.13$ ~~~               \\
    & CDFSE & $\mathbf{4.11\pm0.15}$ & $\mathbf{3.51\pm0.14}$ ~~~ \\
    \bottomrule
  \end{tabular}
  }
\end{table}

As shown in Table \ref{tab:mos}, the results demonstrate our proposed CDFSE method outperforms all three baselines in terms of speaker similarity.
CDFSE gets the best SMOS of $4.11$ for seen speakers and $3.51$ for unseen speakers, and Attentron* performance is relatively better than the two others.
For unseen speakers, the improvement on SMOS of CDFSE is more significant by a gap of over $0.2$, indicating that personal pronunciation characteristics are very helpful to improve the speaker similarity from the sense of listening for zero-shot speaker adaptation.
The MOS results on naturalness of these methods are generally comparable.
CDFSE has a slight decrease in MOS compared with Attentron*, but is still acceptable in terms of naturalness and intelligibility.
This is understandable since frame-level features from reference speech are applied to the TTS decoder output in Attentron*, which helps improve quality and naturalness.

\subsection{Investigation and ablation study}
To investigate the impact of local speaker embeddings with different granularity, we adjust the kernel size of the average pooling layer in the downsample encoders.
In Table \ref{tab:ablation},
the number after `CDFSE-' represents the overall downsampling times
in temporal compared with the reference mel-spectrogram. 
All the models are trained with the same settings as mentioned above. 
We find that some synthesized speeches are poor in intelligibility, which will affect the subjective judgment of similarity.
Therefore, we employ objective evaluations rather than subjective MOS in this part.
To evaluate the intelligibility of synthesized speech, the mispronunciation cases (excluding accents) are marked by listeners and counted.
To evaluate speaker similarity, we employ a speaker verification system \cite{wan2018generalized} to extract the utterance-level speaker vector and calculate the cosine similarity between synthesized speech and ground truth.

\begin{table}[th]
\renewcommand{\arraystretch}{1}
  \vspace{-0.5em}
  \caption{The performance in mispronunciation rate (MPR) and speaker vector cosine similarity (CS) for unseen speaker.}
  \label{tab:ablation}
  \centering
  \resizebox{0.64\columnwidth}{!}{
  \begin{tabular}{l@{}l r r} 
    \toprule
    \multicolumn{2}{c}{\textbf{Model}} &\textbf{MPR} ($\downarrow$) & \textbf{CS} ($\uparrow$)\\
    \midrule
    GSE & & $0.69\%$ & $0.719$ \\
    CLS & & $0.69\%$ & $0.727$ \\
    Attentron* & & $0.69\%$ & $0.737$ \\
    \midrule
    CDFSE-64 & & $0.69\%$ & $0.754$ \\
    CDFSE-16 & & $\mathbf{0.58\%}$ & $\mathbf{0.756}$ \\
    CDFSE-4 & & $11.39\%$ & $0.751$   \\
    CDFSE-1 & & $24.86\%$ & $0.754$   \\
    \midrule
    CDFSE-16 w/o SC & & $1.84\%$ & $0.732$ \\
    \bottomrule
  \end{tabular}
  }
\end{table}

Table \ref{tab:ablation} shows the performance comparison among different granularity models, and the results of three baselines are also presented for reference.
It is observed there exist several mispronunciation cases in all models, which are more likely caused by FastSpeech 2 itself and the training data.
CDFSE-16 gets the lowest mispronunciation rate and the highest speaker vector cosine similarity.
With the decrease of downsampling times, the mispronunciation rate of synthesized speech increases significantly.
That is, the granularity of local speaker embeddings is crucial to the intelligibility and stability of synthesized speech, rather than finer-grained speaker embeddings being better.
This can explain why we use the downsample encoder to extract quasi-phoneme level embedding as stated in 2.1.

Apart from that, we have also employed some ablation studies to demonstrate the effectiveness of each module.
We first remove the explicit supervision of local speaker embedding by excluding speaker classification loss, and this model is denoted as `CDFSE-16 w/o SC' shown in Table \ref{tab:ablation}.
The decline in both two evaluation metrics indicates that introducing speaker information can improve speaker similarity and synthesis stability.
We also remove the explicit supervision of local content embedding by excluding phoneme classification loss, and find it will cause the reference attention module fail.

\subsection{Analysis and discussion}
\begin{figure}[!htb]
	\centering
	\includegraphics[width=0.9\linewidth, ]{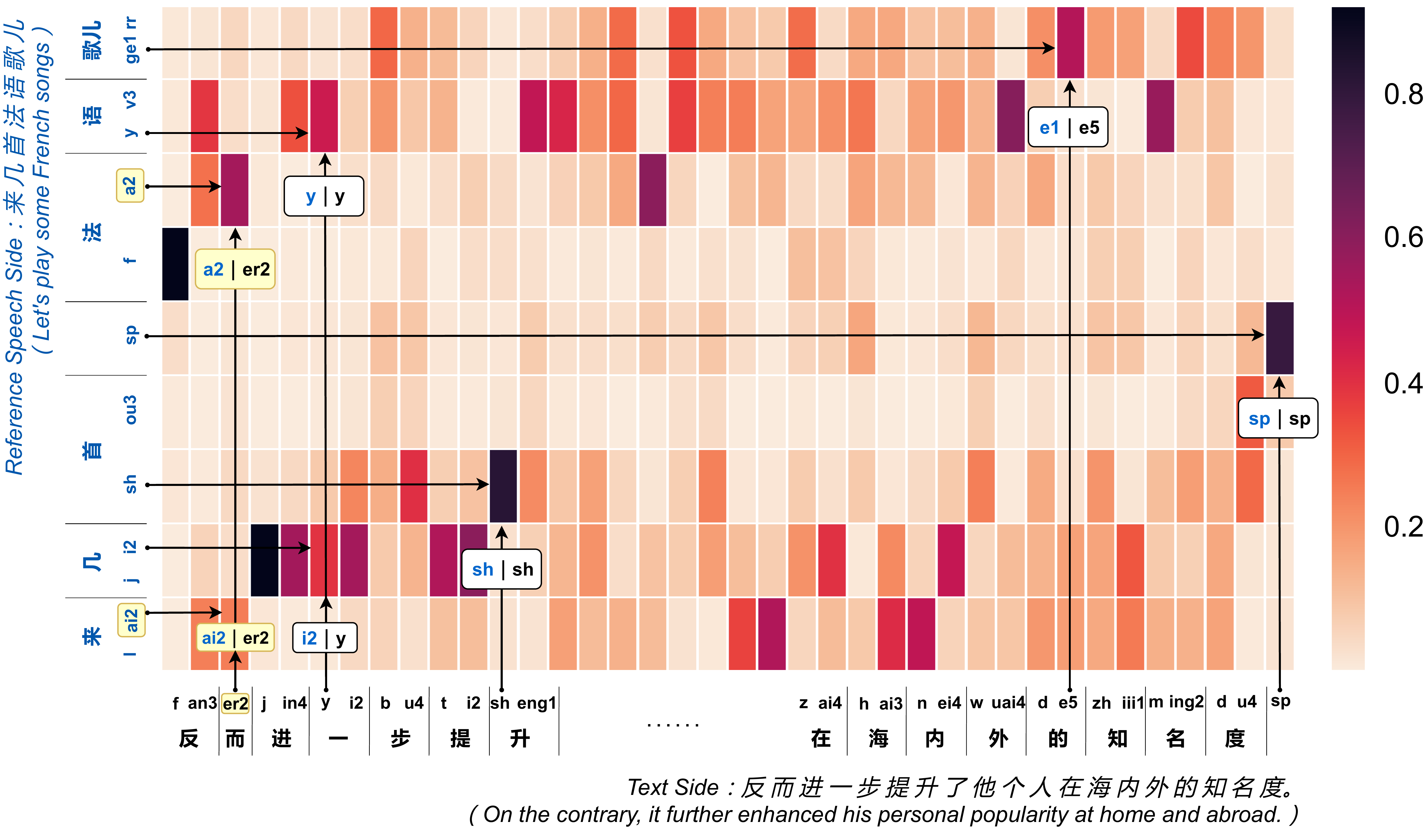}
	\caption{An alignment example in CDFSE. 
	}
	\label{fig:alignment_inrelevant}
\end{figure}

\begin{figure}[htb]
\vspace{-0.8em}
  \begin{minipage}[b]{.48\linewidth}
  \centering
  \includegraphics[width=4.6cm]{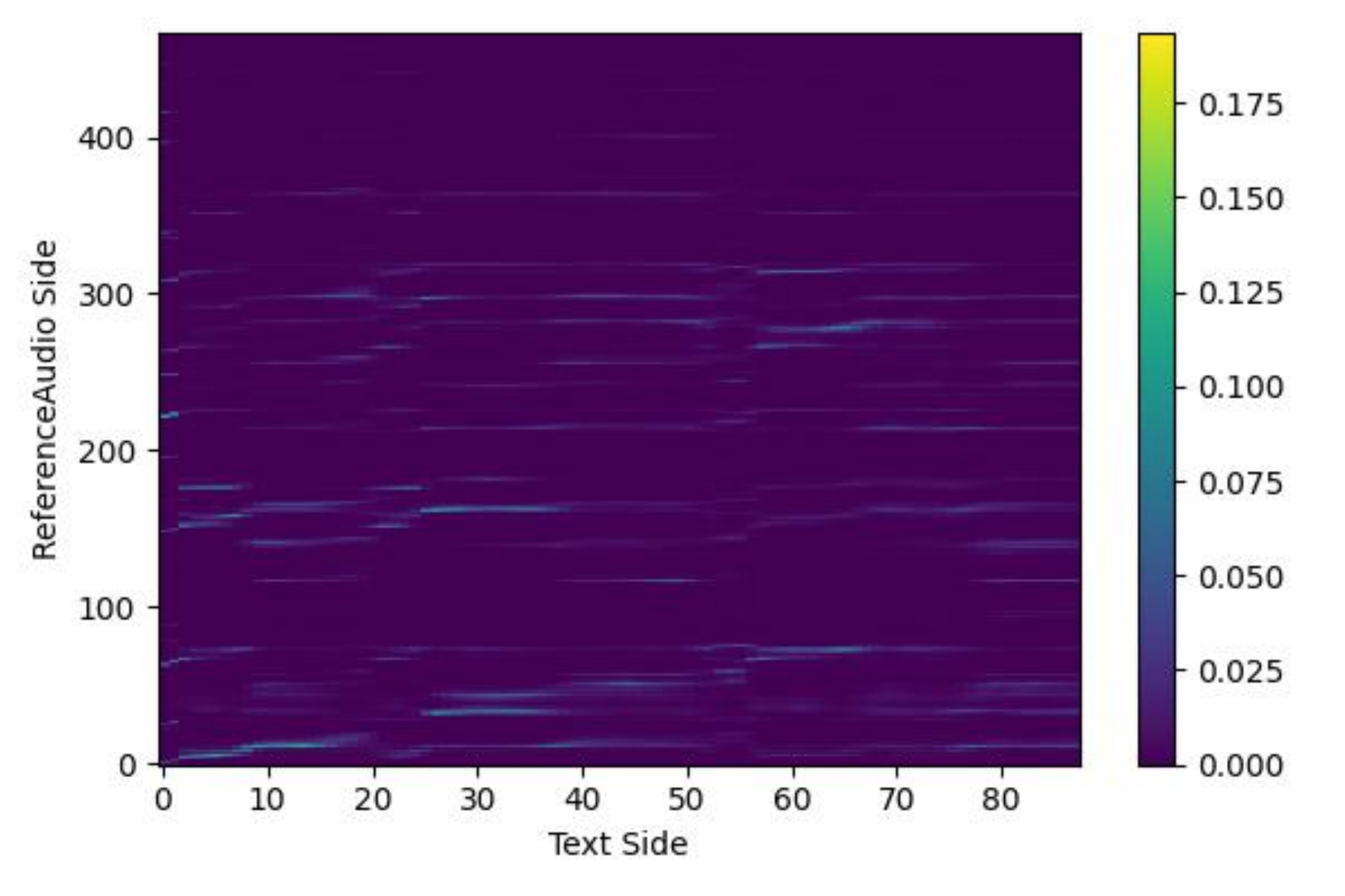} 
  (a) Attentron* 
  \end{minipage}
  \begin{minipage}[b]{.48\linewidth}
  \centering
  \includegraphics[width=4.6cm]{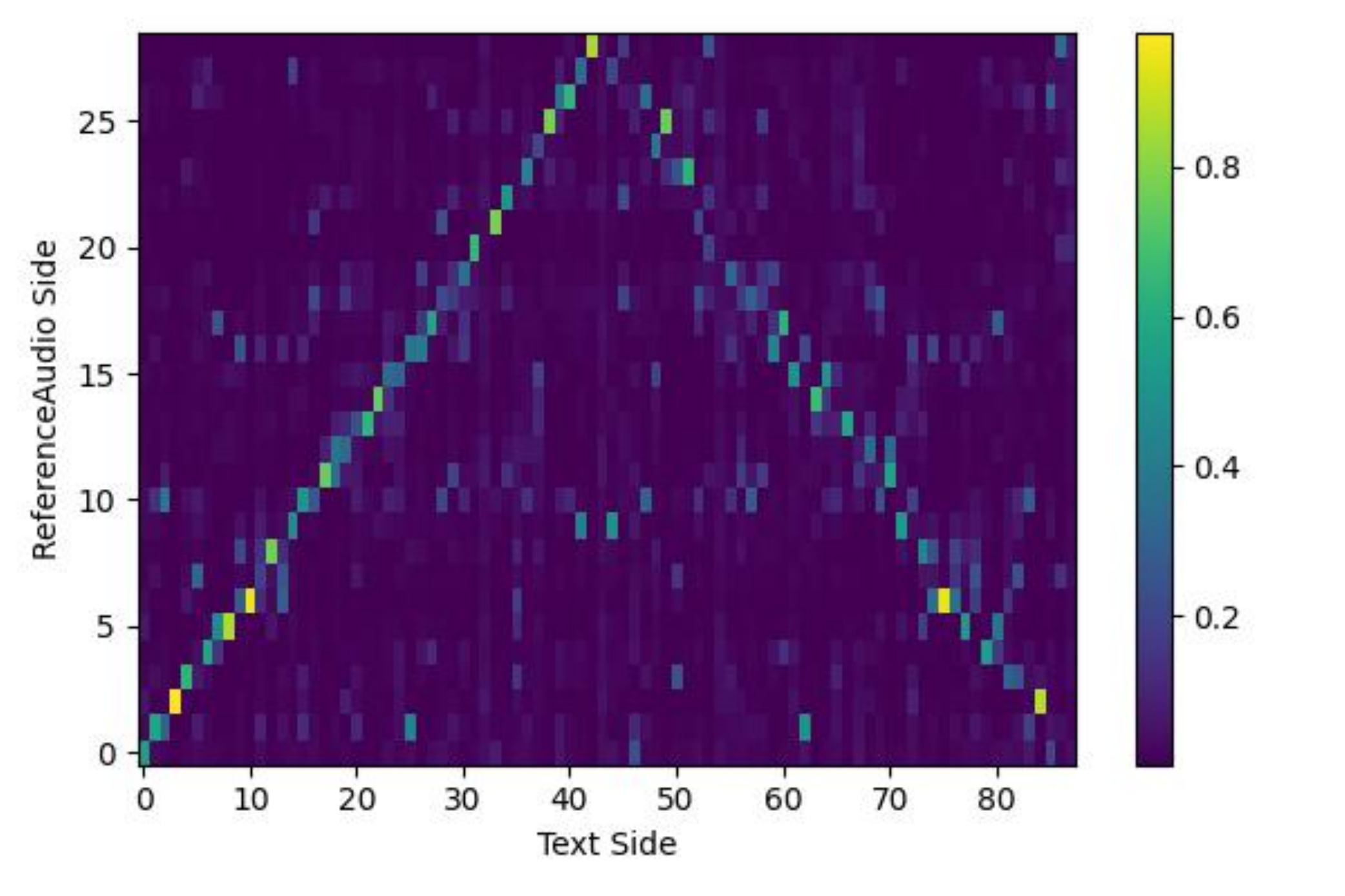}
  (b) CDFSE 
  \end{minipage}
\caption{Alignments between reference speech and input text. The input text is designed to first repeat the content of the reference speech and then reverse it at the Chinese character level.}
\label{fig:alignment}
\end{figure}

\begin{figure}[!htb]
	\centering
	\includegraphics[width=0.63\linewidth, height=0.5\linewidth]{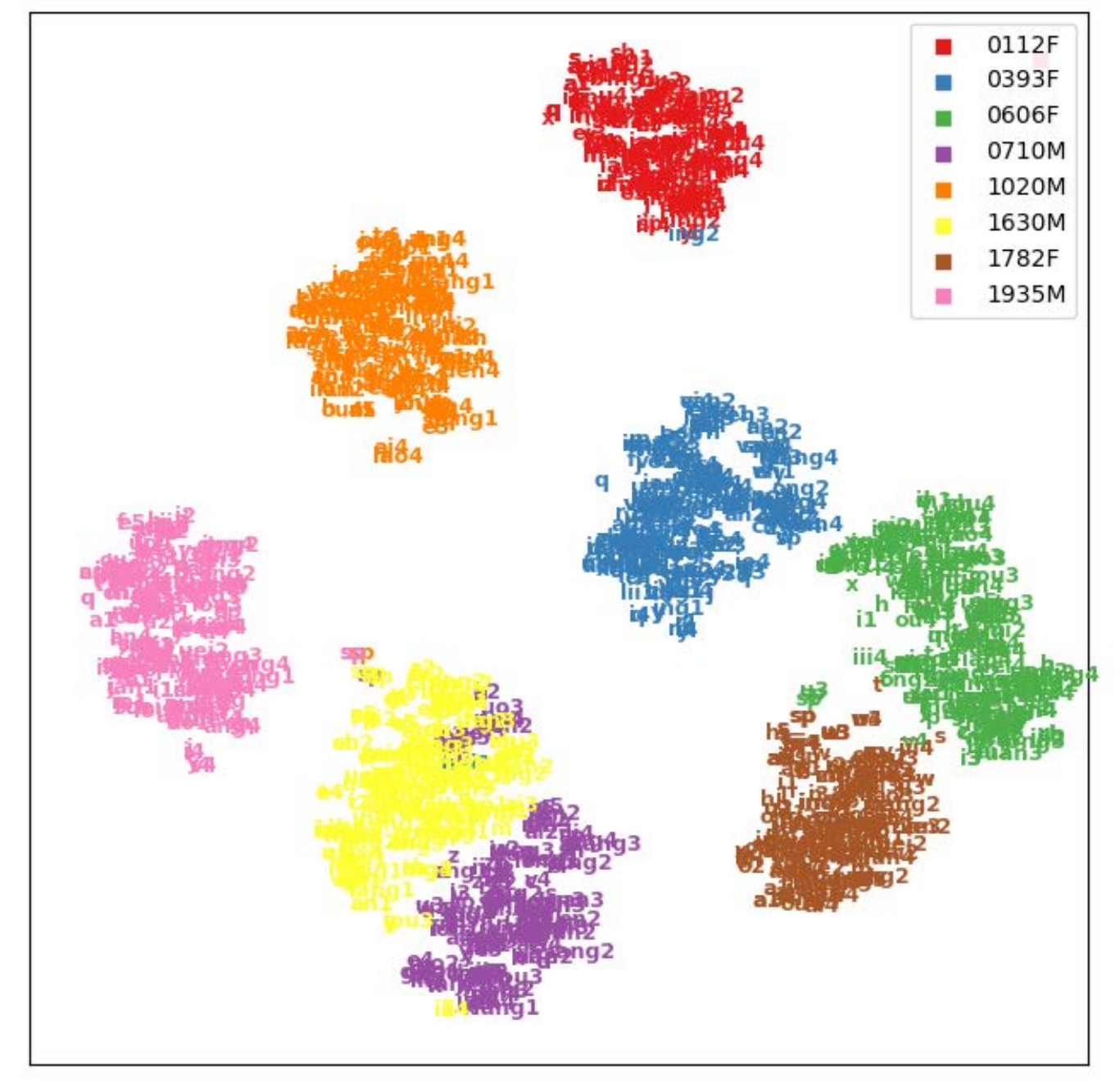}
	\caption{T-SNE visualization of content-dependent fine-grained speaker embeddings. 
	}
	\label{fig:tsne}
	\vspace{-1.3em}
\end{figure}

To clearly present content relevance between reference speech and input text, we plot an alignment example from the reference attention module in CDFSE.
As shown in Fig.\ref{fig:alignment_inrelevant}, when the phoneme in the input text exists in the reference speech, the reference attention tends to focus mainly on the corresponding segment, like ``sh"; 
when the phoneme does not exist, the model will focus on similar segments,
like ``er2" in text similar to ``ai2" and ``a2" in reference speech.
For comparison, another case with specific-designed input text is given, presenting alignments from CDFSE and the attention mechanism in Attentron*.
As shown in Fig.\ref{fig:alignment}, the reference attention module in CDFSE successfully learns the right content alignment (especially, the correct phoneme order within Chinese characters is maintained) between reference speech and text, while Attentron* does not show this ability.

We further visualize the fine-grained speaker embeddings by 2D t-SNE \cite{van2008visualizing}.
As shown in Fig.\ref{fig:tsne}, the fine-grained speaker embeddings of the same speaker tend to group together while exhibiting certain content dependent diversities that capture the local pronunciation variations as stated in 2.2.

\section{Conclusions}
In this paper, we propose content-dependent fine-grained speaker embedding for zero-shot speaker adaptation.
The proposed method can improve the speaker similarity of synthesized speeches, especially for unseen speakers.
Experimental analysis demonstrates that this method has the ability to model personal pronunciation characteristics.

\textbf{Acknowledgement}: This work is supported by National Natural Science Foundation of China (62076144), Tencent AI Lab Rhino-Bird Focused Research Program (JR202143), CCF-Tencent Open Research Fund (RAGR20210122) and Shenzhen Key Laboratory of next generation interactive media innovative technology (ZDSYS20210623092001004).

\newpage

\bibliographystyle{IEEEtran}
\bibliography{references}

\begin{thebibliography}{10}
\providecommand{\url}[1]{#1}
\csname url@samestyle\endcsname
\providecommand{\newblock}{\relax}
\providecommand{\bibinfo}[2]{#2}
\providecommand{\BIBentrySTDinterwordspacing}{\spaceskip=0pt\relax}
\providecommand{\BIBentryALTinterwordstretchfactor}{4}
\providecommand{\BIBentryALTinterwordspacing}{\spaceskip=\fontdimen2\font plus
\BIBentryALTinterwordstretchfactor\fontdimen3\font minus
  \fontdimen4\font\relax}
\providecommand{\BIBforeignlanguage}[2]{{%
\expandafter\ifx\csname l@#1\endcsname\relax
\typeout{** WARNING: IEEEtran.bst: No hyphenation pattern has been}%
\typeout{** loaded for the language `#1'. Using the pattern for}%
\typeout{** the default language instead.}%
\else
\language=\csname l@#1\endcsname
\fi
#2}}
\providecommand{\BIBdecl}{\relax}
\BIBdecl

\bibitem{shen2018natural}
J.~Shen, R.~Pang, R.~J. Weiss, M.~Schuster, N.~Jaitly, Z.~Yang, Z.~Chen,
  Y.~Zhang, Y.~Wang, R.~Skerrv-Ryan \emph{et~al.}, ``Natural tts synthesis by
  conditioning wavenet on mel spectrogram predictions,'' in \emph{2018 IEEE
  International Conference on Acoustics, Speech and Signal Processing
  (ICASSP)}.\hskip 1em plus 0.5em minus 0.4em\relax IEEE, 2018, pp. 4779--4783.

\bibitem{ping2018deep}
W.~Ping, K.~Peng, A.~Gibiansky, S.~O. Arik, A.~Kannan, S.~Narang, J.~Raiman,
  and J.~Miller, ``Deep voice 3: Scaling text-to-speech with convolutional
  sequence learning,'' in \emph{International Conference on Learning
  Representations}, 2018.

\bibitem{ren2020fastspeech}
Y.~Ren, C.~Hu, X.~Tan, T.~Qin, S.~Zhao, Z.~Zhao, and T.-Y. Liu, ``Fastspeech 2:
  Fast and high-quality end-to-end text to speech,'' in \emph{International
  Conference on Learning Representations}, 2021.

\bibitem{zhang2019learning}
Y.~Zhang, R.~J. Weiss, H.~Zen, Y.~Wu, Z.~Chen, R.~Skerry-Ryan, Y.~Jia,
  A.~Rosenberg, and B.~Ramabhadran, ``Learning to speak fluently in a foreign
  language: Multilingual speech synthesis and cross-language voice cloning,''
  in \emph{INTERSPEECH}, 2019, pp. 2080--2084.

\bibitem{chen2020multispeech}
M.~Chen, X.~Tan, Y.~Ren, J.~Xu, H.~Sun, S.~Zhao, T.~Qin, and T.-Y. Liu,
  ``Multispeech: Multi-speaker text to speech with transformer,'' in
  \emph{INTERSPEECH}, 2020, pp. 4024--4028.

\bibitem{tan2021survey}
X.~Tan, T.~Qin, F.~Soong, and T.-Y. Liu, ``A survey on neural speech
  synthesis,'' \emph{arXiv preprint arXiv:2106.15561}, 2021.

\bibitem{arik2018neural}
S.~O. Arik, J.~Chen, K.~Peng, W.~Ping, and Y.~Zhou, ``Neural voice cloning with
  a few samples,'' in \emph{Proceedings of the 32nd International Conference on
  Neural Information Processing Systems}, 2018, pp. 10\,040--10\,050.

\bibitem{chen2021adaspeech}
M.~Chen, X.~Tan, B.~Li, Y.~Liu, T.~Qin, S.~Zhao, and T.-Y. Liu, ``Adaspeech:
  Adaptive text to speech for custom voice,'' in \emph{International Conference
  on Learning Representations}, 2021.

\bibitem{yan2021adaspeech2}
Y.~Yan, X.~Tan, B.~Li, T.~Qin, S.~Zhao, Y.~Shen, and T.-Y. Liu, ``Adaspeech 2:
  Adaptive text to speech with untranscribed data,'' in \emph{2021 IEEE
  International Conference on Acoustics, Speech and Signal Processing
  (ICASSP)}.\hskip 1em plus 0.5em minus 0.4em\relax IEEE, 2021, pp. 6613--6617.

\bibitem{yan2021adaspeech3}
Y.~Yan, X.~Tan, B.~Li, G.~Zhang, T.~Qin, S.~Zhao, Y.~Shen, W.-Q. Zhang, and
  T.-Y. Liu, ``Adaspeech 3: Adaptive text to speech for spontaneous style,'' in
  \emph{INTERSPEECH}, 2021.

\bibitem{jia2018transfer}
Y.~Jia, Y.~Zhang, R.~J. Weiss, Q.~Wang, J.~Shen, F.~Ren, Z.~Chen, P.~Nguyen,
  R.~Pang, I.~L. Moreno \emph{et~al.}, ``Transfer learning from speaker
  verification to multispeaker text-to-speech synthesis,'' in \emph{Proceedings
  of the 32nd International Conference on Neural Information Processing
  Systems}, 2018, pp. 4485--4495.

\bibitem{cooper2020zero}
E.~Cooper, C.-I. Lai, Y.~Yasuda, F.~Fang, X.~Wang, N.~Chen, and J.~Yamagishi,
  ``Zero-shot multi-speaker text-to-speech with state-of-the-art neural speaker
  embeddings,'' in \emph{2020 IEEE International Conference on Acoustics,
  Speech and Signal Processing (ICASSP)}.\hskip 1em plus 0.5em minus
  0.4em\relax IEEE, 2020, pp. 6184--6188.

\bibitem{zhang2021one}
Y.~Zhang, H.~Che, J.~Li, C.~Li, X.~Wang, and Z.~Wang, ``One-shot voice
  conversion based on speaker aware module,'' in \emph{2021 IEEE International
  Conference on Acoustics, Speech and Signal Processing (ICASSP)}.\hskip 1em
  plus 0.5em minus 0.4em\relax IEEE, 2021, pp. 5959--5963.

\bibitem{lu2019one}
H.~Lu, Z.~Wu, D.~Dai, R.~Li, S.~Kang, J.~Jia, and H.~Meng, ``One-shot voice
  conversion with global speaker embeddings,'' in \emph{INTERSPEECH}, 2019, pp.
  669--673.

\bibitem{hsu2018hierarchical}
W.-N. Hsu, Y.~Zhang, R.~J. Weiss, H.~Zen, Y.~Wu, Y.~Wang, Y.~Cao, Y.~Jia,
  Z.~Chen, J.~Shen \emph{et~al.}, ``Hierarchical generative modeling for
  controllable speech synthesis,'' in \emph{International Conference on
  Learning Representations}, 2018.

\bibitem{nguyen2021nvc}
B.~Nguyen and F.~Cardinaux, ``Nvc-net: End-to-end adversarial voice
  conversion,'' \emph{arXiv preprint arXiv:2106.00992}, 2021.

\bibitem{klimkov2019finegrained}
V.~Klimkov, S.~Ronanki, J.~Rohnke, and T.~Drugman, ``Fine-grained robust
  prosody transfer for single-speaker neural text-to-speech,'' \emph{arXiv
  preprint arXiv:1907.02479}, 2019.

\bibitem{lee2019robust}
Y.~Lee and T.~Kim, ``Robust and fine-grained prosody control of end-to-end
  speech synthesis,'' in \emph{2019 IEEE International Conference on Acoustics,
  Speech and Signal Processing (ICASSP)}, 2019, pp. 5911--5915.

\bibitem{li2021towards}
X.~Li, C.~Song, J.~Li, Z.~Wu, J.~Jia, and H.~Meng, ``Towards multi-scale style
  control for expressive speech synthesis,'' in \emph{INTERSPEECH}, 2021.

\bibitem{Fu2019phoneme}
R.~Fu, J.~Tao, Z.~Wen, and Y.~Zheng, ``Phoneme dependent speaker embedding and
  model factorization for multi-speaker speech synthesis and adaptation,'' in
  \emph{2019 IEEE International Conference on Acoustics, Speech and Signal
  Processing (ICASSP)}, 2019, pp. 6930--6934.

\bibitem{choi2020attentron}
S.~Choi, S.~Han, D.~Kim, and S.~Ha, ``Attentron: Few-shot text-to-speech
  utilizing attention-based variable-length embedding,'' in \emph{INTERSPEECH},
  2020, pp. 2007--2011.

\bibitem{ioffe2015batch}
S.~Ioffe and C.~Szegedy, ``Batch normalization: Accelerating deep network
  training by reducing internal covariate shift,'' in \emph{International
  conference on machine learning}.\hskip 1em plus 0.5em minus 0.4em\relax PMLR,
  2015, pp. 448--456.

\bibitem{vaswani2017attention}
A.~Vaswani, N.~Shazeer, N.~Parmar, J.~Uszkoreit, L.~Jones, A.~N. Gomez,
  {\L}.~Kaiser, and I.~Polosukhin, ``Attention is all you need,'' in
  \emph{Advances in Neural Information Processing Systems}, 2017, pp.
  5998--6008.

\bibitem{mcauliffe2017montreal}
M.~McAuliffe, M.~Socolof, S.~Mihuc, M.~Wagner, and M.~Sonderegger, ``Montreal
  forced aligner: Trainable text-speech alignment using kaldi,'' in
  \emph{INTERSPEECH}, 2017, pp. 498--502.

\bibitem{shi2020aishell}
Y.~Shi, H.~Bu, X.~Xu, S.~Zhang, and M.~Li, ``Aishell-3: A multi-speaker
  mandarin tts corpus and the baselines,'' in \emph{INTERSPEECH}, 2021, pp.
  2756--2760.

\bibitem{kong2020hifi}
J.~Kong, J.~Kim, and J.~Bae, ``Hifi-gan: Generative adversarial networks for
  efficient and high fidelity speech synthesis,'' \emph{Advances in Neural
  Information Processing Systems}, vol.~33, 2020.

\bibitem{wan2018generalized}
L.~Wan, Q.~Wang, A.~Papir, and I.~L. Moreno, ``Generalized end-to-end loss for
  speaker verification,'' in \emph{2018 IEEE International Conference on
  Acoustics, Speech and Signal Processing (ICASSP)}.\hskip 1em plus 0.5em minus
  0.4em\relax IEEE, 2018, pp. 4879--4883.

\bibitem{van2008visualizing}
L.~Van~der Maaten and G.~Hinton, ``Visualizing data using t-sne,''
  \emph{Journal of machine learning research}, vol.~9, no.~11, 2008.

\end{thebibliography}

\end{document}